# Ultra-brilliance isolated attosecond gamma-ray light source from nonlinear Compton scattering

Jinqing Yu[1,2], Z. Najmudin[2], Ronghao Hu[1], T. Tajima[3], Haiyang Lu[1,4] & Xueqing Yan[1,4]

**The explosion in attosecond technology has opened the gate to investigating many unexplored areas which require ultrahigh spatial and temporal resolution. In the area of nuclear physics, using gamma-rays with ultrahigh resolution in time and space will help to investigate intra-nuclear dynamics in an unprecedentedly explicit way. However, the generation of ultrahigh brilliance attosecond gamma-ray pulses with current-generation laser facilities has not been reported. In this letter, we propose a novel method to generate high charge (~1nC) attosecond (<200 attosecond) electron bunch by the near-threshold self-injection in a wakefield accelerator. We demonstrate the ability to generate an ultrahigh-brilliance (> $2\times10^{24}$ photons $s^{-1}mm^{-2}mrad^{-2}$ per 0.1%BW) attosecond (<200 attosecond) gamma-ray ($E_{max}$ > 3 MeV) pulse via nonlinear Compton scattering. To the best of our knowledge, this is the first method reported to generate attosecond gamma-ray photon source using current-generation laser. This is the shortest gamma-ray photon and the highest brilliance photon source in MeV range (orders higher than the results[1,2,3] reported). This method can be widely applied for experimental generation of 100 keV to several MeV high brilliance attosecond gamma-ray sources with current ~100 TW laser facilities, which will benefit basic science such as application in ultra-high resolution radiography.**

**Introduction**

X-rays have been widely applied in fundamental research, industry and medicine[4] since their discovery more than a century ago. In fields such as dense object radiotherapy, ultrafast imaging of inner-shell electron dynamics and intra-nuclear dynamics, high quality energetic x-rays (gamma-rays)[1,2,5,6,7] with high brightness and ultra-short duration are particularly desired. Much experimental and theoretical research is directed to making x-ray radiation sources with higher energy, higher brightness, smaller source size and shorter pulse duration[1,2,3,4,8,9,10,11]. However, no method has been reported to generate ultra-high brilliance isolated attosecond light source ranged 100keV to MeV with current laser facilities.

[1]Key Laboratory of HEDP of the Ministry of Education, CAPT, and State Key Laboratory of Nuclear Physics and Technology, Peking University, Beijing 100871, China. [2]The John Adams Institute for Accelerator Science, The Blackett Laboratory, Imperial College London, SW7 2BZ, UK. [3]Department of Physics and Astronomy, University of California, Irvine, CA 92610, USA. [4]Collaborative Innovation Center of Extreme Optics, Shanxi University, Taiyuan, Shanxi, China, 030006. Correspondence and requests for materials should be addressed to X.Q.Y. (x.yan@pku.edu.cn).

Attosecond x-ray sources can be generated via high-order harmonic generation (HHG)[9,10,12], coherent synchrotron emission (CSE)[13,14], the interaction between attosecond electron bunch and ultra-intense few-cycle light[15] and Compton scattering of attosecond gamma-source in the radiation-dominated regime[1]. In the schemes of HHG and CSE, tens of attosecond soft x-ray or extreme-ultraviolet trains are produced experimentsally[12,13]. Low photon energy makes these sources impractical for inner-shell electron dynamics and intra-nuclear dynamics research. Compton scattering, which has been tested by many groups[1,15,16,17] experimentally, is capable of generating high-energy gamma-rays with current laser facilities. However no attosecond gamma-ray source has been reported or theoretically due to the lack of high quality attosecond electron bunch output. In previous studies of strong nonlinear Compton scattering, the generation of attosecond gamma-ray bunch in the radiation-dominated regime[1] with next generation laser facilities[18] is theoretically demonstrated. However, it will be impossible to verify this scheme for some time due to the absence of such high intensity laser facility.

In the Compton scattering scheme, the duration of the radiation is determined by the shorter of either the laser pulse or electron bunch duration. Hence to produce attosecond gamma-ray, two critical points should be overcome. The first is the generation of attosecond electron bunch. In the past few years, several methods were proposed, including laser-solid target interaction[19], laser driven nanoplasma[20] and wakefield acceleration[21,22]. Attosecond electron bunch trains were generated in the first three schemes[19,20,21], and the generation of attosecond electron sheets was proposed by Li et al[22]. All of these methods are not suitable for the generation of isolated micro-spot light source. In order to generate a high density attosecond electron bunch with small transverse size, we report a new method of blow-out wakefield accelerator driven by tightly focused laser beam. Both the effects of an up-ramp density profile and self-focusing are combined to achieve self-balance instantaneous injection of a high quality electron bunch. The generation mechanism of this electron bunch is thus different from that suggested by Li et al[22]. The transverse size of the electron bunch is one order smaller than the electron sheet[22] (tens μm), and is thus better as a micro-spot light source. The second point is the collider after the generation of the electron bunch. We use all-optical Compton mechanism[16] as shown schematically in Figure 1**a**. A critical issue on this point is the collision time to be discussed in detail.

**Theory for the generation of attosecond electron bunch**

In self-injecting laser wakefield acceleration[23,24], the electron bunch density is comparable with the background plasma density, the electron bunch duration is tens of fs[25] as the injection continuously occurs during the acceleration. Electron bunched of high density and short duration can be generated if two critical requirements are both satisfied during the acceleration. They are the increasing of the electron bunch (plasma wave) to high density before injection and the injection of the whole electron wave simultaneously. Ultra-short high density plasma waves can be generated by using plasma of up-ramp density profile[22]. With increasing density of the up-ramp, the bubble size decreases as shown in Figure 1**b**, which inhibits injection[26] and results in the enhancement of the laser intensity. There is a balance point of bubble radius $r_0$ for the given laser intensity and plasma density as given by $k_p r_0 = 2\sqrt{a_0}$ [27], where $k_p$ is the wave number of the plasma wave, $a_0 = 0.85 \times 10^{-9} \lambda(\mu m)\sqrt{I(\text{Wcm}^{-2})}$ and $I$ is the intensity of the laser pulse. The shrinking bubble will turn to be expanded if the size is further reduced after the balance condition, and then the whole plasma wave will inject into the first bubble simultaneously as a result of the bubble expanding[26]. The charge of the injected electrons pushes on trailing electrons forming the sheath, giving then a larger transverse momentum and smaller longitudinal momentum, thus inhibiting subsequent injection. Hence, the electron bunch duration is only determined by the wavelength of the injected plasma wave. The compressed spatial extent of the wake and thus injection event is beneficial for application to Compton scattering. For this, the electron bunch transverse size, which affects the x-ray source size, is required to be on the order of several micro-meters or even smaller.

**Simulation results**

First, a two-dimensional simulation (see Methods) was carried out to study the generation of an attosecond electron bunch in the laser wakefield accelerator with the results shown in Figure 1**b**-1**e**. In this simulation, an under-dense plasma was investigated to test the mechanism of the electron bunch generation and to find the appropriate collision point. The bubble radius $r_0$ is gradually compressed and then increases due to the laser pulse self-focusing and then defocusing during the interaction as shown in figure 1**b**. A high density electron plasma wave is generated due to the phase velocity difference[22] of the first and second bubble. The peak density of the plasma wave is more than two orders higher than background density. After 460 fs, the bubble radius expands and effective electron injection occurs[26]; the whole plasma wave is trapped by the bubble. Figure 1**b**-1**e** show the evolution of the bubble size, bubble breaking in the transverse direction and the self-balance injection. We note the mechanism here is completely different from controllable injection reported by Li[22], in which the plasma profile was specially designed to switch the plasma wave phase velocity. As the bubble rear injects into the bubble as a whole, the vacancy is occupied by the outside electrons with larger transverse momentum. The condition of these electrons is below the requirement of self-injection[26], and subsequent injection is inhibited. Hence, ultra-short isolated electron bunch are generated as shown in Figure 1**e**.

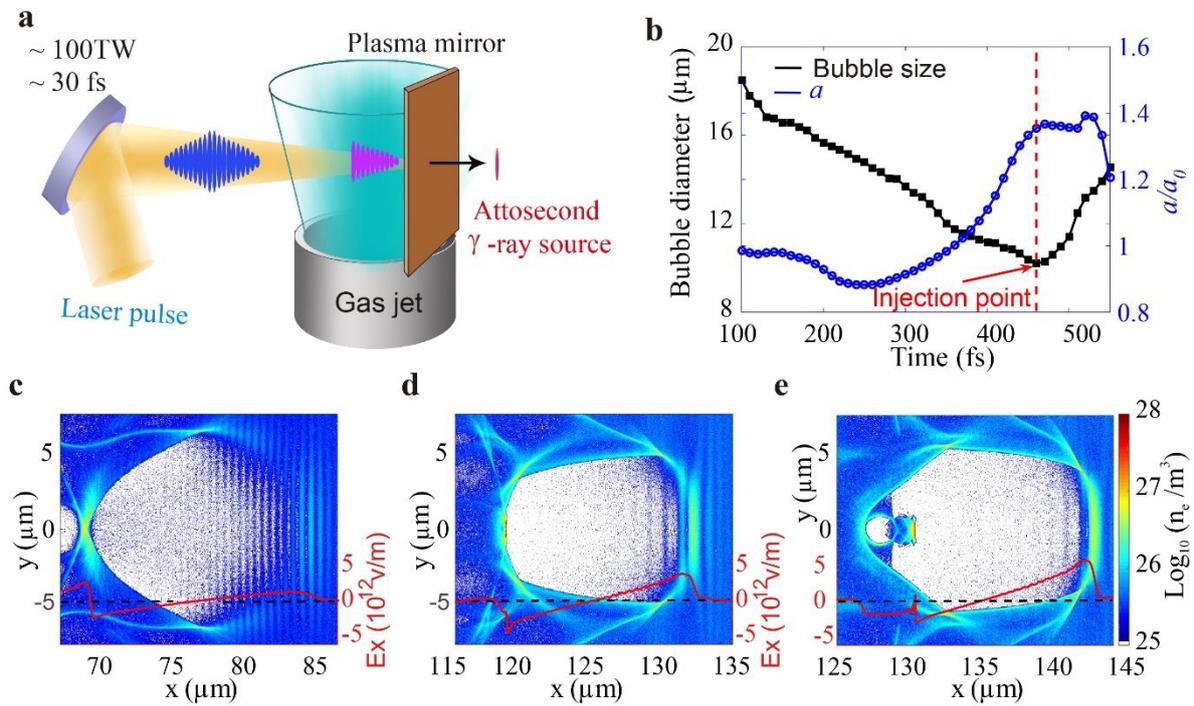

**Figure 1 | Schematic of all-optical gamma-ray light source and attosecond electron bunch generation in a wakefield accelerator. a:** schematic of attosecond gamma-ray source generation from the collision between an attosecond electron bunch. The laser pulse is normally reflected by a plasma mirror and collides with the electron bunch generated in the gas cell, resulting in the production of an isolated ultra-short gamma-ray. The density profile of gas cell target is up-ramp increased from zero, the size of the gas cell can be lager if the density is slower enhanced. **b-e**: 2D PIC simulation results of laser wakefield accelerator in self-injection blow-out regime. **b**: bubble size and laser amplitude *a* vs. simulation time. The bubble diameter is the transverse size of the bubble. In the first 460 fs, the bubble radius $r_0$ is gradually compressed from 11.1 μm to 5.1 μm, while $a_0$ decreases from 8.5 to 7.5 and then increases to 11.6. **c-e**, the electron density and longitudinal electric field on the laser axis (represented by red line) at 300, 460

and 500 fs into the simulation. The plasma wave (in front of the second bubble) injects into the first bubble after 460 fs, resulting in the generation of an attosecond electron bunch. About $4.0 \times 10^9$ electrons are compressed into a bunch of maximum density $n_{e(Max)} \approx 1.2 \times 10^{28}$ m$^{-3}$.

In the blow-out regime of laser wakefield accelerator[23], almost all the electrons are pushed out of the first bubble due to the ponderomotive force of the laser pulse. Figure 2**a** shows the evolution of the electron bunch maximum density after injection. It is found that the density is significantly compressed at the back of the bubble. Before 480 fs, the electron bunch is still surrounded by low density plasma (as shown in Figure 1**d**), although the electron bunch has injected into the bubble. In that case, the bunch transverse oscillation is still inhibited and the density modulation is at a relatively low level. After 480 fs, the electrons are further accelerated into the bubble and transverse oscillation become significant, resulting in the electron density enhancement after 490 fs as shown in Figure 2**a**. The charge of the electron bunch is about 1.0 nC, the duration is less than 200 attosecond (Figure 2**b**), a number density above $1.2 \times 10^{28}$ m$^{-3}$ can be achieved and the source radius is about 1 μm. The acceleration distance is only tens μm, hence the kinetic energy of the electron is only tens MeV as show in Figure 2**c**. The electron bunch has a relative large spread angle of 65 mrad due to the large charge in the bunch as show in Figure 2**d**.

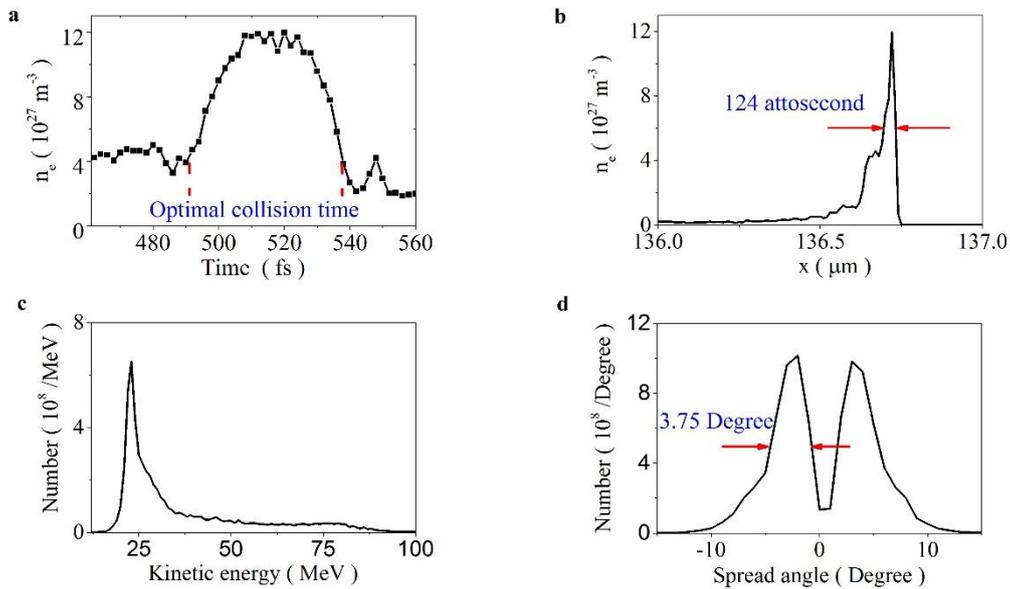

**Figure 2 | Results of attosecond electron bunch. a**: the electron bunch maximum density history between 460 and 560 fs from the same simulation as fig. 1. The electron bunch reaches higher density from 492 to 538 fs, which corresponds to the optimal collision time. **b-d**: results of the electron bunch at 520 fs into the

simulation. **B**: electron number density on the laser axis, the shortest duration of the electron bunch is about 124 attosecond. **c**: the energy spectrum of the electron bunch peaks at 23 MeV. **d:** the angle distribution of the electron bunch, the divergence angle is about 3.75 degree.

Second, a simulation was performed including the reflection mirror (see methods) to investigate the photon radiation. To make sure that the electron bunch has the highest density and smallest source size during the collision, we place the plasma mirror at 141 μm. After the laser-electron collision, a high dense attosecond photon bunch is generated. It is found that the photon bunch duration is about 230 attosecond which is comparable with the electron bunch duration. While the duration of the gamma-ray (> 100 keV) is about 67 attosecond as shown in Figure 3**a** &**c**. The total photon number $2.2 \times 10^{10}$ is more than 3 times higher than the total electron number, which indicates that multiple photon emission occurs per electron during the nonlinear Compton scattering. From Figure 3**b**, we can see that photon with higher kinetic energy has smaller spread angle, and the divergence angle is about 9.5 degrees for all the photons, 6.5 degrees for the photons > 100 keV and 3.5 degrees for the photons > 1 MeV. The photon maximum energy is more than 3 MeV as shown in Figure 3**d**.

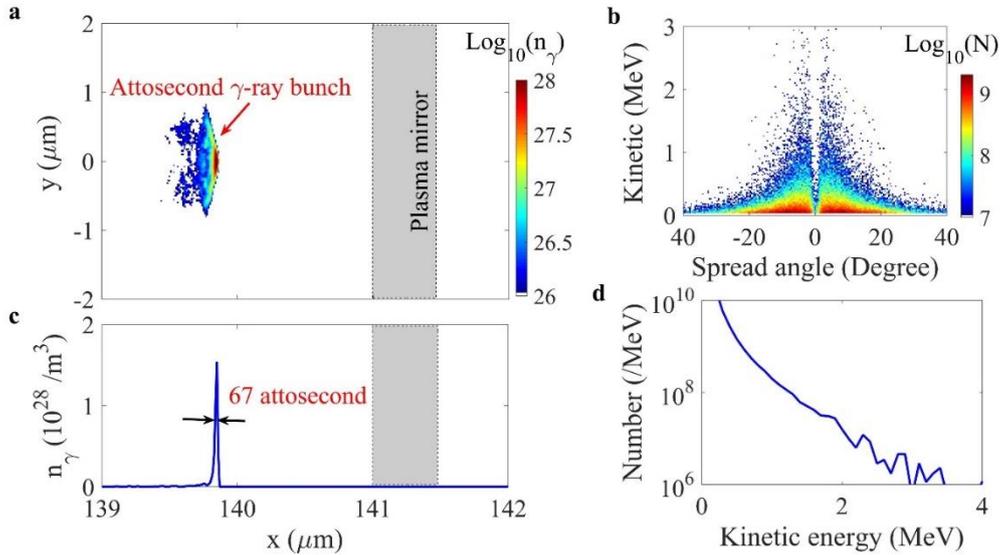

**Figure 3 | Results of attosecond x-ray bunch.** The effective collision occurs from 513 fs to 527 fs, resulting in the generation of $N_p \approx 2.2 \times 10^{10}$ photons. The total gamma photon number (>100keV) is about $3.3 \times 10^9$, with $1.9 \times 10^8$ photons above 1 MeV. **a:** gamma photon density distribution, **b:** divergence angle $\theta_p = \tan^{-1}(p_\perp/p_x)$, as a function of photon kinetic energy, **c:** gamma photon density on the laser axis at 530 fs, peak photon density $n_p$ is more than $1.5 \times 10^{28}$ m$^{-3}$, the photon source size is

about 1 μm and the bunch duration is about 67 attosecond FWHM, **d**: energy spectrum of the photon bunch.

Here, we consider photons with kinetic energy above 100 keV to be gamma-ray photon. Figure 4**a** shows the anglular resolution of the attosecond gamma-ray bunch at different times. It is found that the gamma pulse is very stable after the laser-electron collision. Inside the source width (~1 μm), the gamma photon number is about $3.0 \times 10^6$ in 0.1% bandwidth at 100 keV and $9.0 \times 10^5$ in the 0.1% bandwidth at 1 MeV. Hence, the brilliance is about $2.8 \times 10^{24}$ $s^{-1}mrad^{-2}mm^{-2}$ per 0.1%BW at 100 keV and about $2.9 \times 10^{24}$ $s^{-1}mrad^{-2}mm^{-2}$ per 0.1%BW at 1 MeV. We carry out a test simulation in the full 3D condition, from which we find that the physical features are almost the same as illustrated from 2D simulations. An ultra-short gamma-ray source is also generated in the 3D simulation as show in Figure 4**b**. We also carried out more simulations with larger plasma size up to several millimeters but slower enhancement of the density profile. In all the simulations, high brilliance attosecond gamma-ray bunch can be generated.

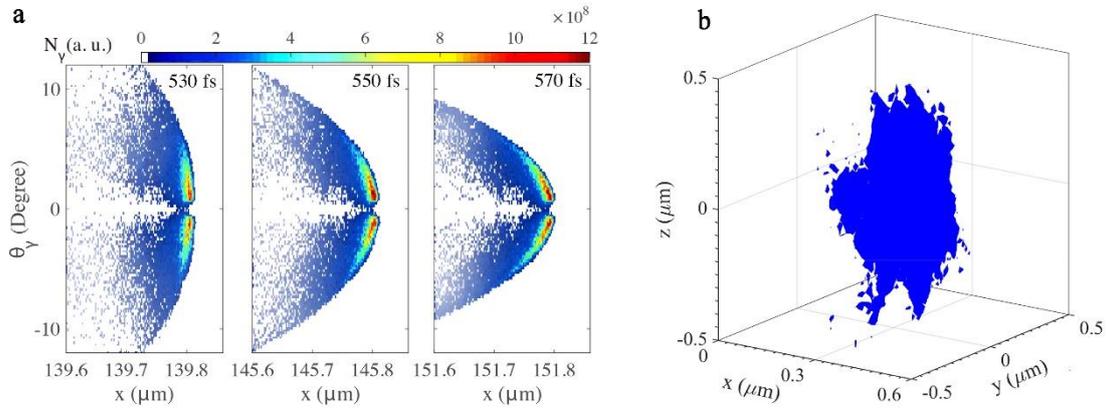

**Figure 4 | Results of attosecond gamma-ray bunch. a:** the angular distribution of the gamma-ray bunch at different times from 2D-PIC simulation. Only photons with kinetic energy > 100 keV are plotted in the figure, **b:** the gamma-ray density from 3D-PIC simulation at 540 fs, the iso-value $n_\gamma > 2.0 \times 10^{27}$ $m^{-3}$ is plotted in the figure and the maximum density of the gamma-ray bunch is more than $n_{\gamma\ (Max)} \approx 1.0 \times 10^{28}$ $m^{-3}$.

Conclusion, we theoretically propose a novel method to generate a high brilliance isolated attosecond gamma-ray with current-generation laser. The feasibility is demonstrated by numerical simulations. In the integrated simulation, the generation of isolated attosecond electron bunch and the collision time are overcame and high brilliance isolated attosecond gamma-ray bunch is generated.

**Methods**

We use the two-dimensional relativistic particle-in-cell code EPOCH2D[28] to model the generation of attosecond electron bunch and gamma-ray. In the PIC code, the quantum synchrotron radiation[29] is included. In the laser field, the photon emission probability of the accelerated particle is governed by a parameter named optical depth $\tau_e$. The position, momentum and $\tau_e$ of the emitting particle are updated every computational time-step. More detail on the method to calculate the radiation numerically can be found in the work of Duclous et al[29]. Simulations are carried out on Super Computation Center in Max Planck Institute. We use a linearly polarized laser pulse with a transverse spatial Gaussian and $\sin^2$ temporal profile. The laser pulse normally irradiates a gas cell plasma from the left. The laser duration full width at half maximum (FWHM) is 30 fs and the focal spot size is 8.0 μm, giving an intensity $1.0 \times 10^{20}$ W/cm$^2$. Such laser parameters can be easily achieved by many 100TW laser facilities around the world. A hydrogen plasma is used with density linearly increasing in the longitudinal direction from $x=0$ with a function as $n_e(x) = 2 \times 10^{25} x/(60\mu m)$ m$^{-3}$ and a fully ionized aluminum film is modeled as a plasma mirror. The thickness of the plasma mirror is 500 nm to save computing resource, but could be much thicker in the experiment. The hydrogen plasma and the plasma mirror are located from 0 to 141 μm and 141 to 141.5 μm, respectively. The gas cell could be much thicker if the density is slower enhanced. The simulation time is 600 fs. In the 2D-PIC simulations, the simulation box size is 180 μm in the $x$ direction and 30 μm in $y$, and cell size is $d_x \times d_y$=10 nm×20 nm. We put 2 particles per species in each cell of the hydrogen plasma, while 1300 electrons and 100 ions are placed in each cell for the plasma mirror. In the 3D simulation, we used a moving window to model the same physics process as observed from 2D simulation. The cell size is $d_x \times d_y \times d_z$=10 nm×80 nm×80 nm. As in the 2D simulation, the hydrogen plasma had, 2 particles per species in each cell, while for the plasma mirror, 130 electrons and 10 ions are placed in each cell.

**Acknowledgements**

The work has been supported by the National Basic Research Program of China (Grant No.2013CBA01502, 11475010), NSFC (Grant Nos.11535001), National Grand Instrument Project (2012YQ030142), the Projects (2016M600007, 2017T100009) funded by China Postdoctoral Science Foundation. J. Q. Y. wants to thank the fruitful suggestions from Dr. Wenjun Ma, Z. Gong and H. X. Chang. The PIC code Epoch was in part funded by the UK EPSRC grants EP/G054950/1. Our simulations were carried out in Max Planck Computing and Data Facility and CX1 at Imperial College London.

**Author contributions**

J. Q. Y. and X. Q. Y. designed this research and drafted the manuscript. J. Q. Y carried out the simulations. All the authors contributed in finalizing the manuscript.

**Competing financial interests**

The authors declare no competing financial interests.